\documentclass[twocolumn,pre]{revtex4}

\usepackage{graphicx}
\usepackage{amsmath,amssymb}
\usepackage{calrsfs}

\newcommand\mat{\mathbf}
\newcommand\e{\mathrm{e}}

\renewcommand\vec{\mathbf}

\newcommand\etal{\textit{et~al.}}
\newcommand\half{\tfrac12}

\newcommand{\omegain}{\omega_\textrm{in}}
\newcommand{\omegaout}{\omega_\textrm{out}}
\renewcommand{\min}{m_\textrm{in}}
\newcommand{\mout}{m_\textrm{out}}

\begin{document}

\title{Community detection in networks:\\
Modularity optimization and maximum likelihood are equivalent}

\author{M. E. J. Newman}
\affiliation{Department of Physics and Center for the Study of Complex
  Systems, University of Michigan, Ann Arbor, MI 48109}

\begin{abstract}
We demonstrate an exact equivalence between two widely used methods of community detection in networks, the method of modularity maximization in its generalized form which incorporates a resolution parameter controlling the size of the communities discovered, and the method of maximum likelihood applied to the special case of the stochastic block model known as the planted partition model, in which all communities in a network are assumed to have statistically similar properties.  Among other things, this equivalence provides a mathematically principled derivation of the modularity function, clarifies the conditions and assumptions of its use, and gives an explicit formula for the optimal value of the resolution parameter.
\end{abstract}

\maketitle

\section{Introduction}
\label{sec:intro}
Community detection, sometimes called network clustering, is the division of the nodes of an observed network into groups such that connections are dense within groups but sparser between them~\cite{POM09,Fortunato10,Newman12}.  Not all networks support such divisions, but many do, and the existence of good divisions is often taken as a hint of underlying semantic structure or possible mechanisms of network formation, making community detection a useful tool for interpreting network data.

The development of methods or algorithms to perform community detection on empirical networks has been a popular pursuit among researchers in physics, mathematics, statistics, and computer science---a tremendous number of such algorithms have been published in the last decade or so~\cite{POM09,Fortunato10,Newman12,CGP11}.  In this paper we study two of the most popular and widely used methods for community detection in simple undirected networks, the method of modularity maximization and the method of maximum likelihood as applied to the stochastic block model.  Building on previous work by ourselves and others~\cite{Newman13a,Newman13b,ZM14}, we show that, different though they at first appear, these two methods are in fact exactly equivalent, for appropriate choices of models and parameters, which we specify.  This sheds light in particular on the modularity maximization method, which is generally motivated with heuristic arguments~\cite{NG04,Newman04a,RB06a} (although there are some rigorous results~\cite{Brandes07,BC09}).  Our results provide a rigorous derivation for the modularity and demonstrate that modularity maximization is optimal under appropriate conditions, but also highlight the method's limitations.  In particular, we show that modularity maximization effectively assumes that communities in a network are statistically similar, and it is not guaranteed to give good results for networks where this is not true.

\section{Community detection}
We begin by describing the two methods of community detection that we study, in their most widely accepted forms, beginning with the method of modularity maximization.

\subsection{Modularity maximization}
Modularity maximization is perhaps the most widely used method for community detection for networks.  It operates by defining a benefit function, called the modularity, that measures the quality of divisions of a network into communities.  One optimizes this benefit function over possible divisions of the network of interest to find the one that gives the highest score, taking this to be the definitive division of the network.  Since the number of possible divisions of a network is exponentially large, we normally cannot perform the optimization exhaustively, so we turn instead to approximate optimization methods, of which many have been tried, including greedy algorithms~\cite{Newman04a,CNM04}, extremal optimization~\cite{DA05}, spectral relaxation~\cite{Newman06b}, genetic algorithms~\cite{SYHF09}, simulated annealing~\cite{GSA04,MAD05}, and belief propagation~\cite{ZM14}.  The popular Louvain algorithm for community detection~\cite{BGLL08}, which is built into a number of network analysis software packages, uses a multiscale modularity optimization scheme and is one of the fastest community detection methods in practice.

The definition of the modularity function is straightforward~\cite{NG04}.  We desire a benefit function which, given a network and a candidate division of that network into groups, returns a score that is larger if the division is a ``good'' one and smaller if it is ``bad.''  The heuristic notion used to define the modularity is that a good division is one that places most of the edges of a network within groups and only a few of them between groups.

Let us represent our network by its adjacency matrix.  For an undirected network of $n$ nodes the adjacency matrix~$\mat{A}$ is the real symmetric $n\times n$ matrix with elements $A_{ij}=1$ if there is an edge between nodes $i$ and~$j$ and 0 otherwise.  Further, let us consider a division of the network into $q$ nonoverlapping groups, numbered (in any order) from 1 to~$q$, and let us denote by $g_i$ the number of the group to which node~$i$ is assigned.  Thus the complete vector~$\vec{g}$ of group assignments specifies the division of the network.  Then the number of edges that fall within groups, for this particular division, is equal to $\half \sum_{ij} A_{ij} \delta_{g_ig_j}$, where $\delta_{ij}$ is the Kronecker delta and the leading factor of a half prevents double counting of~edges.

The number of in-group edges alone, however, is not a good measure of the quality of a division, since it can be trivially maximized by putting all the nodes in one of the $q$~groups and none in any of the others.  This would put 100\% of edges inside groups but clearly doesn't constitute a useful division of the network.  Instead, therefore, modularity measures not just the number of edges within groups but the difference between that number and the expected number of such edges, were edges placed at random within the network.

Suppose we take our observed network and randomize the positions of its edges. We keep the total number of edges the same but we reposition them between the nodes at random, in a manner to be determined shortly.  And suppose that, following this randomization, the probability that nodes $i$ and $j$ are connected by an edge is~$P_{ij}$.  Then the expected number of edges within groups, post-randomization, is $\half \sum_{ij} P_{ij} \delta_{g_ig_j}$.  The modularity is then proportional to the actual minus expected number of edges thus:
\begin{align}
Q &= {1\over m} \biggl( \half \sum_{ij} A_{ij} \delta_{g_ig_j}
                        - \half \sum_{ij} P_{ij} \delta_{g_ig_j} \biggr)
     \nonumber\\
  &= {1\over2m} \sum_{ij} \bigl( A_{ij} - P_{ij} \bigr) \delta_{g_ig_j}
\label{eq:genmod}
\end{align}
where $m$ is the total number of edges in the network and is included here by convention only---it makes~$Q$ equal to a \emph{fraction} of edges rather than an absolute number, which makes modularity values more easily comparable between networks of different sizes.  For the purposes of maximizing the modularity, which is our main concern here, the factor of $m$ makes no difference at all.  The position of the maximum does not depend on overall constant factors.

Note that if we now put all nodes in the same group, then $\delta_{g_ig_j}=1$ for all $i,j$ and
\begin{equation}
Q = {1\over2m} \sum_{ij} \bigl( A_{ij} - P_{ij} \bigr) = 0,
\end{equation}
since, as we have said, the number of edges in the network is held constant during randomization, and hence $\sum_{ij} P_{ij} = \sum_{ij} A_{ij} = 2m$.  Thus we no longer get a high modularity score for putting all nodes in a single group together.  The maximum of modularity occurs for some other (nontrivial) division of the nodes, which we take to be the best division of the network.  This is the method of modularity maximization.

It remains to determine what $P_{ij}$ is.  The value depends on the particular scheme we use to randomize the positions of the edges.  The simplest scheme would be just to reposition the edges uniformly at random, every position having the same probability as every other.  For a network of $n$ nodes there are ${n\choose2}$ places to put an edge, and hence the probability of filling any of them with one of the $m$ edges is
\begin{equation}
P_{ij} = {m\over{n\choose2}}.
\label{eq:ndcmod}
\end{equation}
(Technically this is the expected number of edges not the probability, but normally $m\ll{n\choose2}$ so that probability and expected number are essentially the same.)

In practice, however, this choice does not work very well because it fails to respect the degrees of the nodes in the network.  The probabilities of connections between nodes depend strongly on the total number of connections nodes have---their degrees---with nodes of high degree being much more likely to be connected than nodes of low degree.  For reasons that will become clear in this paper, it is important to include this effect in the definition of modularity if things are to work correctly.

Instead, therefore, we consider a constrained randomization of the edges in the network in which we preserve the degree of every node, but otherwise position the edges at random.  This kind of randomization is well known in the study of networks: it gives rise to the random graph ensemble known as the configuration model~\cite{MR95,NSW01}.  After randomization, the probability of connection between two nodes is equal to
\begin{equation}
P_{ij} = {k_i k_j\over2m},
\label{eq:dcmod}
\end{equation}
where $k_i = \sum_j A_{ij}$ is the degree of node~$i$ and $m$ is once again the number of edges in the network.  (Again, this is technically the expected number of edges, but the probability and expected number are essentially the same.)

This is the choice that is most commonly used in the definition of the modularity.  With this choice the modularity is given by
\begin{equation}
Q = {1\over2m} \sum_{ij} \biggl( A_{ij} - {k_i k_j\over2m} \biggr)
    \delta_{g_ig_j},
\label{eq:modularity}
\end{equation}
which is the form in which it is most commonly written.

There is a further twist, however, because even this definition does not always work well.  As shown by Fortunato and Barth\'el\'emy~\cite{FB07}, community detection by modularity maximization using the definition of~\eqref{eq:modularity}, while it works in many situations, has one specific shortcoming: it is unable to find community structure in large networks with many small communities.  In particular, if the number of communities in a network is greater than about~$\sqrt{2m}$, then the maximum modularity will not correspond to the correct division.  The maximum will instead tend to combine communities into larger groups and fail to resolve the smallest divisions in the network.

To address this problem, Reichardt and Bornholdt~\cite{RB06a} proposed a generalized modularity function
\begin{equation}
Q(\gamma) = {1\over2m} \sum_{ij} \biggl( A_{ij} - \gamma {k_i k_j\over2m}
             \biggr) \delta_{g_ig_j}.
\label{eq:generalized}
\end{equation}
When the parameter~$\gamma=1$, this is the same as the traditional modularity of Eq.~\eqref{eq:modularity}, but other choices allow us to vary the relative weight given to the observed and randomized edge terms.  If one places more weight on the observed edge term (by making $\gamma$ smaller), the maximum modularity division favors, and the method therefore tends to find, larger communities.  If one places more weight on the randomized edge term (larger~$\gamma$), the method finds smaller communities.  There has not previously been any fundamental theory dictating what value of $\gamma$ one should use, but this is one of the questions on which we will shed light in this paper.

\subsection{Statistical inference}
\label{sec:inference}
The other method of community detection we consider is the method of statistical inference, as applied to the stochastic block model and its variants.  With this method, one fits a generative model of a network to observed network data, and the parameters of the fit tell us about the structure of the network in much the same way that fitting a straight line through a set of data points tells us about their slope.

The model most commonly used in this context is the so-called stochastic block model, which is a random graph model of a network with community structure~\cite{HLL83,NS01,BC09}.  One takes some number~$n$ of nodes, initially without any edges, and divides them into~$q$ groups in some way, with $g_i$ being the group to which node~$i$ is assigned, as previously.  Then one places edges between nodes independently at random, with the probability, which we denote~$\omega_{rs}$, of an edge between a particular pair of nodes depending on the groups $r$ and~$s$ to which the nodes belong.  Thus there is a symmetric $q\times q$ matrix of parameters~$\omega_{rs}$ which determine the probabilities of edges within and between every pair of groups.  If the diagonal elements~$\omega_{rr}$ of this matrix are larger than the off-diagonal elements, then networks generated by the model have a higher probability of edges within groups than between them and hence have traditional community structure.

In fact, the stochastic block model is often studied in a slightly different formulation in which one places not just a single edge between any pair of nodes but a Poisson distributed number of edges with mean~$\omega_{rs}$.  Thus $\omega_{rs}$ is the expected number, rather than the probability, of edges between nodes in groups~$r$ and~$s$, and the networks generated by the model can in principle have multiedges, meaning there can be more than one edge between the same pair of nodes.  Moreover, one typically also allows the network to contain self-edges, edges that connect a node to itself, which are also Poisson distributed in number, with mean $\half\omega_{rr}$ for a node in group~$r$.  (The factor of half is included solely because it makes the algebra simpler.)  The inclusion of multiedges and self-edges in the model is unrealistic in the sense that most network data encountered in the real world contain neither.  However, most real-world networks are also very sparse, meaning that only a tiny fraction of all possible edges that could exist actually do, and hence the values of the edge probabilities $\omega_{rs}$ are very small.  In this situation, the density of multiedges and self-edges in the network will itself be small and the Poisson version of the model is virtually indistinguishable from the first (Bernoulli) version defined above.  At the same time, the Poisson version is technically easier to handle than the Bernoulli version.  In this paper we use the Poisson version.  Similar results can be derived for the Bernoulli version, but the formulas are more complicated and the end result is little different.

The definition of the model above is in terms of its use to generate networks.  As applied to community detection, however, the model is used in the ``reverse'' direction to infer structure by fitting it to data.  In this context, one hypothesizes that an observed network, with adjacency matrix~$\mat{A}$, was generated from the stochastic block model, and attempts to work out what values of the model parameters must have been used in the generation.  The parameters in this case are the edge probabilities~$\omega_{rs}$ and the group memberships~$g_i$.

Given particular values of the parameters we can write down the probability, or likelihood, that the observed network was generated from the block model thus:
\begin{equation}
P(\mat{A}|\mat{\Omega},\vec{g})
  = \prod_i {(\half\omega_{g_ig_i})^{A_{ii}/2}\over (\half A_{ii})!}
    \e^{-\omega_{g_ig_i}/2}
    \prod_{i<j} {\omega_{g_ig_j}^{A_{ij}}\over A_{ij}!} \e^{-\omega_{g_ig_j}},
\end{equation}
where $\mat{\Omega}$ denotes the complete matrix of values~$\omega_{rs}$ and we have adopted the standard convention that a self-edge is represented by a diagonal adjacency matrix element~$A_{ii}=2$ (and not 1 as one might at first imagine).

The position of the maximum of this quantity with respect to $\mat{\Omega}$ and $\vec{g}$ tells us the values of the parameters most likely to have generated the observed network.  Typically we are interested only in the group assignments~$\vec{g}$, which tell us how the network divides into groups.  The probabilities~$\mat{\Omega}$ are usually uninteresting and can be discarded once the likelihood has been maximized.

Alternatively (and usually more conveniently), we can maximize the logarithm of the likelihood:
\begin{align}
\log &P(\mat{A}|\mat{\Omega},\vec{g}) \nonumber\\
  &= \sum_i \bigl[ \half A_{ii} \log \bigl( \half\omega_{g_ig_j} \bigr)
     - \half\omega_{g_ig_j} - \log \bigl( \half A_{ij}! \bigr) \bigr] \nonumber\\
  &\qquad{} + \sum_{i<j} \bigl( A_{ij} \log \omega_{g_ig_j} - \omega_{g_ig_j}
    - \log A_{ij}! \bigr).
\end{align}
The terms $\half A_{ii} \log \half$, $\log(\half A_{ij}!)$, and $\log A_{ij}!$ are all independent of the parameters and do not affect the position of the maximum, so they can be ignored, and the log-likelihood then simplifies to
\begin{equation}
\log P(\mat{A}|\mat{\Omega},\vec{g}) = \half \sum_{ij} \bigl( A_{ij} \log \omega_{g_ig_j} - \omega_{g_ig_j} \bigr).
\end{equation}
The optimal division of the network into communities is then given by maximizing this quantity with respect to both $\vec{g}$ and~$\mat{\Omega}$.

\subsection{Degree-corrected block model}
\label{sec:dcsbm}
As with the modularity, however, this is not the whole story.  This approach fares poorly when applied to most real-world networks because it doesn't respect the node degrees in the network.  The stochastic block model as described here (in either Bernoulli or Poisson versions) generates networks that have a Poisson degree distribution, which is very different from the broad distributions seen in empirical networks.  This means that, typically, the model does not fit observed networks well for any choice of parameter values.  It's as if one were trying to fit a straight line through an inherently curved set of data points.  Even the best fit of such a line will not be a good fit.  There are no good fits when the model you are fitting is simply wrong.

The conventional solution to this problem, in the present situation, is to use a slightly different model, the degree-corrected block model~\cite{KN11a}, which can fit networks with any degree distribution.  In this model the nodes are again assigned to groups~$g_i$ and edges placed independently at random between them, but now the expected number of edges between nodes $i$ and $j$ is $(k_i k_j/2m) \omega_{rs}$ (where as before $r,s$ are the groups to which the nodes belong and $k_i,k_j$ are the degrees) or a half that number for self-edges.  The factor of $2m$ in the denominator is optional, but convenient since, as we have said, $k_i k_j/2m$ is the probability of an edge in the configuration model and hence, with this definition, $\omega_{rs}$~quantifies the probability of edges relative to the configuration model.

Following the same line of reasoning as before, and again neglecting constants that have no effect on the position of the likelihood maximum, the log-likelihood for this model is
\begin{equation}
\log P(\mat{A}|\mat{\Omega},\vec{g}) = \half \sum_{ij} \biggl( A_{ij} \log \omega_{g_ig_j} - {k_i k_j\over 2m} \omega_{g_ig_j} \biggr).
\label{eq:dcsbm}
\end{equation}

Community detection now involves the maximization of this quantity with respect to the parameters $\mat{\Omega},\vec{g}$ to find the best fit of the model to the observed network.  This maximization can be achieved in a number of ways.  As with the modularity, there are too many possible group assignments~$\vec{g}$ to maximize exhaustively on any but the smallest of networks, but researchers have successfully applied a variety of approximate methods, including label switching algorithms~\cite{BC09}, Kernighan--Lin style greedy algorithms~\cite{KL70,KN11a}, spectral methods~\cite{Newman13b}, Monte Carlo~\cite{NS01,Peixoto14b}, and belief propagation~\cite{DKMZ11a,Yan14}.

\section{The planted partition model
and modularity maximization}
\label{sec:equivalence}
We now come to the central result of this paper, the equivalence of modularity maximization to a particular case of the maximum likelihood method described above.  We previously discussed a version of this equivalence in the context of work on spectral algorithms~\cite{Newman13b,Newman13a} and it has also been discussed by Zhang and Moore~\cite{ZM14} in the context of work on finite-temperature ensembles of graph partitions.  Building on these works, our purpose in this paper is to make explicit the exact equivalence of the two approaches and investigate some of its consequences.

The planted partition model~\cite{CK01,McSherry01} is a special case of the stochastic block model in which the parameters~$\omega_{rs}$ describing the community structure take only two different values:
\begin{equation}
\omega_{rs} = \biggl\lbrace \begin{array}{ll}
                \omegain  & \qquad\text{if $r=s$,} \\
                \omegaout & \qquad\text{if $r\ne s$.}
              \end{array}
\label{eq:ppmodel}
\end{equation}
This is a less flexible model than the full stochastic block model.  It effectively assumes that all communities in the network are similar in the sense of having the same in-group and between-group connection rates.  Nonetheless, for networks that do have this property, fits to the model should recover the community structure accurately, and indeed it has been proved that such fits are optimal in that case~\cite{DKMZ11a,Massoulie14,MNS15}.

In practice, if one wanted to apply the planted partition model, one should in almost all cases use a degree-corrected version of the kind described in Section~\ref{sec:dcsbm}.  Let us explore the form of the log-likelihood, Eq.~\eqref{eq:dcsbm}, for such a model.  Following~\cite{Newman13a,Newman13b} we note that Eq.~\eqref{eq:ppmodel} implies that
\begin{align}
\omega_{rs}      &= (\omegain - \omegaout) \delta_{rs} + \omegaout, \\
\log \omega_{rs} &= (\log\omegain - \log\omegaout) \delta_{rs}
                      + \log\omegaout,
\end{align}
where $\delta_{rs}$ is the Kronecker delta, as previously.  Substituting these forms into Eq.~\eqref{eq:dcsbm}, we find the log-likelihood for the degree-corrected planted partition model to be
\begin{align}
\log P(\mat{A}|\mat{\Omega},\vec{g})
  &= \half \sum_{ij} A_{ij} \biggl[ \delta_{g_ig_j} \log {\omegain\over\omegaout}
     + \log \omegaout \biggr] \nonumber\\
  &\qquad{} - \half \sum_{ij} {k_i k_j\over 2m} \bigl[ (\omegain-\omegaout)
     \delta_{g_ig_j} + \omegaout \bigr] \nonumber\\
  &\hspace{-5.5em}{} = \half \log {\omegain\over\omegaout}
     \sum_{ij} \biggl( A_{ij} -
     {(\omegain-\omegaout)\over(\log\omegain-\log\omegaout)}\,{k_i k_j\over 2m}
     \biggr) \delta_{g_ig_j} \nonumber\\
  &+ m \bigl( \omegaout + \log\omegaout \bigr) \nonumber\\
  &\hspace{-5.5em}{} = B\,{1\over2m}
     \sum_{ij} \biggl( A_{ij} - \gamma {k_i k_j\over 2m} \biggr) \delta_{g_ig_j}
     + C,
\label{eq:main}
\end{align}
where $B$ and $C$ are constants that depend on $\omegain$ and $\omegaout$ but not on~$\vec{g}$, and
\begin{equation}
\gamma = {\omegain-\omegaout\over\log\omegain-\log\omegaout}.
\label{eq:equivgamma}
\end{equation}
We have also made use of $\sum_{ij} A_{ij} = \sum_i k_i = 2m$ in the second equality of~\eqref{eq:main}.

To perform community detection, one would now maximize this expression with respect to both the group assignments~$g_i$ and the parameters~$\omegain$ and~$\omegaout$.  But suppose for a moment that we already know the correct values of $\omegain$ and~$\omegaout$, leaving us only to maximize with respect to the group assignments.  Comparing Eq.~\eqref{eq:main} with Eq.~\eqref{eq:generalized}, we see that, apart from overall constants, \eqref{eq:main}~is precisely the generalized modularity~$Q(\gamma)$, and hence the likelihood and the modularity have their maxima with respect to the~$g_i$ in the same place.  Thus community detection by maximization of the likelihood for the planted partition model with known values of $\omegain$ and~$\omegaout$ is equivalent to maximizing the generalized modularity for the appropriate value of~$\gamma$, given by Eq.~\eqref{eq:equivgamma}.  (We leave it as a exercise for the reader to show that a similar equivalence applies---with the same value of~$\gamma$---between maximizing the likelihood for the non-degree-corrected stochastic block model and the modularity when one makes the choice~\eqref{eq:ndcmod} for~$P_{ij}$.)

Among other things, this result tells us what the correct value of the resolution parameter~$\gamma$ is for the generalized modularity, an issue that has hitherto been undecided.  The correct value is given by Eq.~\eqref{eq:equivgamma}.  An immediate corollary is that in most cases the conventional choice $\gamma=1$, corresponding to the original, non-generalized modularity function of Eq.~\eqref{eq:modularity}, is not correct.

Unfortunately, however, we cannot normally employ Eq.~\eqref{eq:equivgamma} directly to calculate~$\gamma$, since we do not know the values of the parameters $\omegain$ and~$\omegaout$.  We present one possible solution to this problem in Section~\ref{sec:gamma}, but for the moment let us proceed under the assumption that we know the correct value of~$\gamma$.

The equivalence between modularity maximization and maximum-likelihood methods has a number of immediate implications.  First of all, it provides a derivation of the modularity that is more rigorous and principled than the usual heuristic arguments: modularity maximization (with the correct choice of~$\gamma$) is equivalent to fitting a network to a degree-corrected version of the planted partition model using the method of maximum likelihood.  It also explains why the standard degree-dependent choice, Eq.~\eqref{eq:dcmod}, for the definition of the modularity is better than the uniform choice of Eq.~\eqref{eq:ndcmod}.  It is for the same reason that the degree-corrected block model is the correct choice for the analysis of most real-world networks: the uniform choice effectively assumes a network with a Poisson degree distribution, which is a poor approximation to most empirical networks.  The degree-dependent choice, by contrast, fits networks of any degree distribution.

The equivalence of modularity and maximum-likelihood methods also implies that modularity maximization is a consistent method of community detection, i.e.,~that under suitable conditions it will correctly and without bias find community structure where present.  This follows because maximum-likelihood fits to stochastic block models are also known to be consistent: when the maximum-likelihood method is applied to networks that were themselves generated from the same block model (either the traditional or degree-corrected model), it will return correct assignments of nodes to groups in the limit of large node degrees~\cite{BC09}.  The consistency of modularity maximization has been demonstrated previously by other means~\cite{ZLZ11,ZM14}, but the equivalence with likelihood maximization makes the physical intuition behind it particularly clear.

A further point of interest is that while the value of $\gamma$ in Eq.~\eqref{eq:equivgamma} is always positive, regardless of the values of $\omegain$ and~$\omegaout$, the value of the constant $B = m\log(\omegain/\omegaout)$ in Eq.~\eqref{eq:main} changes sign depending on which of $\omegain$ and $\omegaout$ is larger.  This means that maximization of the likelihood becomes equivalent to \emph{minimization} of the modularity when $\omegaout>\omegain$, i.e.,~when the network has so-called disassortative structure, in which connections are more common between groups than within them.  The minimization of modularity to find such structure has been proposed previously on heuristic grounds~\cite{Newman06c}, but the derivation here gives a rigorous foundation for the procedure.

On the other hand, the equivalence of maximum likelihood and maximum modularity methods also reveals some hidden assumptions and limitations of the modularity.  The planted partition model, with its assumption, Eq.~\eqref{eq:ppmodel}, that the edge parameters~$\omega_{rs}$ take the same values for every community, is less powerful than the full stochastic block model and modularity maximization is similarly less powerful as a result.  In effect, modularity maximization assumes all communities in a network to be statistically similar.  This may be a good assumption in some networks, but there are certainly examples where it is not, and we would expect the modularity maximization method to perform less well in such cases than more general methods.

Some variants of the maximum-likelihood method also include additional parameters that allow for heterogeneous group sizes, but the version used here, to which modularity maximization is equivalent, includes no such embellishments, meaning that modularity maximization implicitly favors groups of uniform size, which could also hurt performance.

\subsection{Value of the resolution parameter}
\label{sec:gamma}
A drawback of the equivalence we have demonstrated is that it applies only when we use the correct value of the resolution parameter~$\gamma$, which normally we do not know.  One can, however, make an empirical estimate of the value of~$\gamma$ using an iterative scheme as follows.

First, one makes an initial guess about the value of~$\gamma$.  This guess need not be particularly accurate: $\gamma=1$~usually works fine.  Then one finds the communities in the network by modularity maximization using this choice.  This gives us some set of assignments~$g_i$ of nodes to groups---likely not optimal---from which we can then make an estimate of the parameters $\omegain$ and~$\omegaout$ by noting that the expected total number of in-group edges~$\min$ in the (degree-corrected) planted partition model is
\begin{equation}
\min = \half \sum_r \sum_{ij} {k_i k_j\over2m} \omega_{rr} \delta_{g_i,r} \delta_{g_j,r}
     = {\omegain\over4m} \sum_r \kappa_r^2,
\end{equation}
where $\kappa_r = \sum_i k_i \delta_{g_i,r}$ is the sum of the degrees of the nodes in group~$r$.  Hence we can estimate $\omegain$ from
\begin{equation}
\omegain = {2\min\over\sum_r \kappa_r^2/2m},
\end{equation}
using the observed value of~$\min$ as an estimate of the expected value.  Similarly for $\omegaout$ we have
\begin{equation}
\omegaout = {2\mout\over\sum_{r\ne s} \kappa_r\kappa_s/2m}
          = {2m-2\min\over 2m - \sum_r \kappa_r^2/2m},
\end{equation}
where $\mout$ is the number of edges running between distinct groups.

\begin{figure}
\begin{center}
\includegraphics[width=7.8cm]{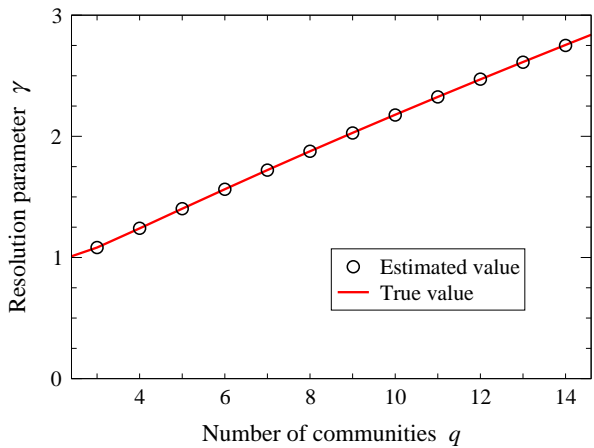}
\end{center}
\caption{Resolution parameter~$\gamma$, estimated using the method described here, for a set of synthetic networks with varying numbers of communities.  The networks were generated using the standard (non-degree-corrected) planted partition model with~$q$ equally sized groups of 250 nodes each and parameters~$\omegain$ and~$\omegaout$ chosen so that each node has an average of 16 connections within its own group and 8 to every other group.  Modularity was maximized using simulated annealing.  The circles represent the estimated values of $\gamma$ and the solid line represents the true values calculated from Eq.~\eqref{eq:equivgamma}.}
\label{fig:sbm}
\end{figure}

Using these estimates of $\omegain$ and~$\omegaout$ we can now calculate a new value of $\gamma$ from Eq.~\eqref{eq:equivgamma}.  Then we repeat the process, maximizing the modularity and recalculating~$\omegain$, $\omegaout$, and~$\gamma$ until we achieve convergence.  The consistency of modularity maximization, mentioned earlier, implies that this procedure should converge to the correct value of~$\gamma$ (and the correct community structure) for sufficiently dense networks that actually are generated from the planted partition model.  For all other networks (meaning, in practice, for all real-world applications of the method) we have no formal guarantees of correctness or convergence, though the same is also true of all other methods of community detection, including, but not limited to, community detection by statistical inference.

\begin{table}
\begin{center}
\setlength{\tabcolsep}{4pt}
\begin{tabular}{lrrrc}
Network & $n$ & $m$ & $q$ & $\gamma$ \\
\hline
Karate club                             & 34   & 78    & 2  & 0.78 \\
Dolphin social network                  & 62   & 159   & 2  & 0.59 \\
Political blogs                         & 1225 & 16780 & 2  & 0.67 \\
Books about politics                    & 105  & 441   & 2  & 0.59 \\
Characters from \textit{Les Miserables} & 77   & 254   & 6  & 1.36 \\
American college football               & 115  & 614   & 11 & 2.27 \\
Jazz collaborations                     & 198  & 2742  & 16 & 1.19 \\
Email messages                          & 1133 & 5451  & 26 & 3.63 \\
\end{tabular}
\end{center}
\caption{Number of nodes~$n$, number of edges~$m$, number of communities~$q$, and estimated value of the resolution parameter~$\gamma$ for a range of networks studied in the previous literature.  The networks are the karate club network of Zachary~\cite{Zachary77}, the dolphin social network of Lusseau~\etal~\cite{Lusseau03a}, the network of political weblogs studied by Adamic and Glance~\cite{AG05}, the network of books about politics studied by Krebs (unpublished, but see for instance~\cite{Newman06b}), the network of interactions between fictional characters in the novel \textit{Les Miserables} by Victor Hugo~\cite{NG04}, the network of regular season games between Division I-A college football teams in the year 2000~\cite{NG04}, the network of collaborations between jazz musicians studied by Gleiser and Danon~\cite{GD03}, and the network of email messages between university students of Ebel~\etal~\cite{EMB02}.  The value of~$q$ used for each network is the generally accepted one, except for the last two networks, for which there does not appear to be a consensus.  For these two networks we estimate the number of communities using the inference method of~\cite{NR16}.}
\label{tab:networks}
\end{table}

One might imagine that this would not be a very efficient method for calculating~$\gamma$: it requires repeated maximization of different modularity functions until the correct value of~$\gamma$ is reached.  In practice, however, we have found that it converges very quickly.  In most cases we have examined, $\gamma$~is calculated to within a few percent after just one iteration, and in no case have we found a need for more than ten iterations, so the method may in fact be quite serviceable. Figure~\ref{fig:sbm} shows an example application to a set of artificially generated (``synthetic'') networks for which the true value of~$\gamma$ is known and, as the figure shows, the algorithm is able to determine that value accurately in every case.  Table~\ref{tab:networks} gives values of $\gamma$ calculated using the algorithm for a number of real-world networks that have been used as test cases in previous community detection studies.

The values of $\gamma$ vary in size, but there is an overall trend towards larger values in networks with larger numbers of communities, both among the synthetic networks and the real ones.  This is perhaps not unexpected given that the resolution parameter~$\gamma$ was originally introduced precisely in order to deal with networks with larger numbers of communities.  Recall that larger values of~$\gamma$, and specifically values larger than the traditional value of~1, are needed in networks where the number of communities exceeds the resolution limit at~$\sqrt{2m}$.  None of the networks studied here approach this limit, but nonetheless we should not find it surprising that the larger values of~$q$ in both Fig.~\ref{fig:sbm} and Table~\ref{tab:networks} are best treated using values $\gamma>1$.

Whether the algorithm given here is in fact a useful one in practice is a debatable point.  As we have shown, it does no more than the likelihood maximization method, and the latter in principle gives better results, since it does not assume that all groups are statistically similar.  Modularity maximization does have the advantage of being less nonlinear than maximum-likelihood methods, which allows for some faster algorithms such as spectral~\cite{Newman06b,Newman13b} and multiscale~\cite{BGLL08} algorithms.  Still, the results derived here are primarily of interest not because of the algorithms they suggest, but because of the light they shed on the strengths and weaknesses of modularity maximization.

\section{Conclusions}
We have shown that modularity maximization is a special case of the maximum likelihood method of community detection, as applied to the degree-corrected planted partition model.  The equivalence between the two approaches highlights some weaknesses of the modularity maximization method: the method assumes all communities to have statistically similar properties, which may not be the case, and it also contains an undetermined parameter~$\gamma$.  Most often this parameter is assumed to take the value~1, but we have shown that this assumption is normally not correct and given an explicit formula for the correct value, along with a simple algorithm for computing it on observed networks.

\begin{acknowledgments}
The author thanks Aaron Clauset, Travis Martin, Cris Moore, and Cosma Shalizi for useful comments.  This research was supported in part by the US National Science Foundation under grants DMS--1107796 and DMS--1407207.
\end{acknowledgments}

\end{document}